\documentclass[12pt]{article}
\usepackage{amsmath,amssymb}
\usepackage{graphicx}
\usepackage[pdftex]{hyperref}
\DeclareGraphicsExtensions{.eps,.bmp,.wmf,.jpeg,.pdf}
\numberwithin{equation}{section}
\def\be{\begin{equation}}
\def\ee{\end{equation}}

\def\bea{\begin{eqnarray}}
\def\eea{\end{eqnarray}}
\title{Dark energy from scalar field with Gauss Bonnet and non-minimal kinetic coupling}
\author{L.N. Granda\thanks{ngranda@univalle.edu.co} \\ {\small\it Departamento de Fisica, Universidad del Valle}\\{\small\it A.A. 25360, Cali, Colombia}} 

\date{}
\begin{document}
\maketitle

\begin{abstract}
\noindent  We study a model of scalar field with a general non-minimal kinetic coupling to itself and to the curvature, and additional coupling to the Gauss Bonnet 4-dimensional invariant. The model presents rich cosmological dynamics and some of its solutions are analyzed. A variety of scalar fields and potentials giving rise to power-law expansion have been found. The dynamical equation of state is studied for two cases, with and without free kinetic term . In both cases phenomenologically acceptable solutions have been found. Some solutions describe essentially dark energy behavior, and and some solutions contain the decelerated and accelerated phases.\\ 

\noindent PACS 98.80.-k, 95.36+x, 04.50.kd
\end{abstract}

\section{Introduction}
\noindent 
The late time acceleration of the universe is one of the most important problems of modern cosmology, which is supported by astrophysical data from distant Ia supernovae observations \cite{riess}, \cite{perlmutter}, \cite{kowalski}, \cite{hicken}, cosmic microwave background anisotropy \cite{komatsu}, and large scale galaxy surveys \cite{percival}. The interpretation of astrophysical observations indicates that this accelerated expansion is due to some kind of  negative-pressure form of matter known as dark energy (\cite{copeland}, \cite{sahnii}, \cite{padmanabhan}). The combined analysis of cosmological observations also suggests that the universe is spatially flat, and  consists of about $\sim 1/3$ of dark matter, and $\sim 2/3$ of homogeneously distributed dark energy with negative pressure. The dark energy may consist of cosmological constant, conventionally associated with the energy of the vacuum \cite{peebles}, \cite{padmana1}, or alternatively, could came from a dynamical varying scalar field at late times which also account for the missing energy density in the universe. A widely explored scalar field models are quintessence \cite{RP}, \cite{wett}, \cite{copeland97}, tachyon \cite{pad}, phantom \cite{caldwell}, K-essence \cite{stein3},\cite{chiba}, and dilaton \cite{gasperini} (for a review see \cite{copeland}). The scalar fields are allowed from several theories in particle physics and in multidimensional gravity, like Kaluza-Klein theory, String Theory or Supergravity, in which the scalar field appears in a natural way. Another alternative to the explanation
of the DE is represented by the scalar-tensor theories, which contain a direct coupling of the scalar field to the curvature, providing in principle a mechanism to
evade the coincidence problem, and naturally allowing (in some cases) the crossing of the phantom barrier \cite{peri}, \cite{maeda}. From a pure geometrical point of view, the modified gravity theories, which are generalizations of the general relativity, have been widely considered to describe the early-time inflation and late-time acceleration, without the introduction of any other dark component, and represent an important alternative to explain the dark energy (for review see \cite{sergei11} and references therein).\\
In the present work, we consider a model with non-minimal coupling to gravity \cite{amendola2, capozziello1, capozziello2}, and with additional Gauss Bonnet coupling, specifically we focus in a scalar field model with kinetic term non-minimally coupled to gravity and to itself \cite{granda,granda1,granda2}, with a new term containing the Gauss Bonnet (GB) 4-dimensional invariant coupled to the scalar field, with the coupling given by an arbitrary function of the field. 

Despite the fact that the GB term is topologically invariant in four dimensions, and hence, by itself does not contribute to the equations of motion, nevertheless it affects cosmological dynamics when it is coupled to a dynamically evolving scalar field. Besides this, if the GB term is coupled to the scalar field through arbitrary function $f(\phi)$, then this is the unique quadratic combination of the Riemann curvature tensor, that does not increase the differential order of the equations of motion (i.e. GB produces only terms which are second derivatives of the metric in the field equations). Therefore, the coupled GB term which preserves the theory ghost free, seems as a natural generalization of the scalar field with non-minimal kinetic coupling to curvature, which under certain condition is a second order scalar tensor theory. 
Both couplings my have origin in the low energy limit of higher dimensional theories. Thus, the kinetic coupling appears as part of the Weyl anomaly in $N=4$ conformal supergravity \cite{tseytlin, odintsov2}, while the GB coupling arises naturally in the leading order of the $\alpha'$ expansion of string theory \cite{callan}, \cite{bento}. Besides that, the kinetic couplings to curvature are also present as quantum corrections to Brans-Dicke theory \cite{elizalde} and in different frames in modified gravity \cite{sergei11}.\\
Some late time cosmological aspects of scalar field model with derivative couplings to curvature have been considered in \cite{sushkov}, \cite{saridakis}, \cite{gao}. On the other hand, the GB invariant coupled to scalar field have been extensively studied. In \cite{sergei12} the GB correction was proposed to study the dynamics of dark energy, where it was found that quintessence or phantom phase may occur in the late time universe. Accelerating cosmologies with GB correction in four and higher dimensions have been discussed in \cite{tsujikawa}, \cite{leith}, \cite{maartens}. The modified GB theory applied to dark energy have been suggested in \cite{sergei14}, and different aspects of the modified GB model applied to late time acceleration, have been considered among others, in \cite{sergei11}, \cite{sergei15}, \cite{carter}, \cite{tretyakov}.\\
All these studies demonstrate that it is quite plausible that some scalar-tensor couplings predicted by the fundamental theory may become important at current, low-curvature
universe. The aim of the present paper is the study of the possible effects of the Gauss-Bonnet coupling combined with the kinetic coupling, 
on dark energy cosmologies, in order to clarify whether they could be compatible with current phenomenology.

\section{Field Equations}
Let us start with the  action, for scalar field kinetic terms non-minimally coupled to curvature and coupled to Gauss Bonnet (GB) curvature
\be\label{eq1}
\begin{aligned}
S=&\int d^{4}x\sqrt{-g}\Big[\frac{1}{16\pi G} R-\frac{1}{2}\partial_{\mu}\phi\partial^{\mu}\phi-\frac{1}{2} \xi R \left(F_1(\phi)\partial_{\mu}\phi\partial^{\mu}\phi\right) -\\ 
&\frac{1}{2} \eta R_{\mu\nu}\left(F_1(\phi)\partial^{\mu}\phi\partial^{\nu}\phi\right) - V(\phi)+F_2(\phi){\cal G}\Big]+S_m.
\end{aligned}
\ee
\noindent where {\cal G} is the 4-dimensional GB invariant ${\cal G}=R^2-4R_{\mu\nu}R^{\mu\nu}+R_{\mu\nu\rho\sigma}R^{\mu\nu\rho\sigma}$, $S_m$ is the dark matter action which describes a fluid with barotropic equation of state. The dimensionality of the coupling constants $\xi$ and $\eta$ depends on the type of function $F_1(\phi)$, and the coupling $F_2(\phi)$ is dimensionless. Besides the couplings of curvatures with kinetic terms, one may expect that the presence of GB coupling term may be relevant for the explanation of dark energy phenomena.\\
Taking the variation of action (\ref{eq1}) with respect to the metric, we obtain a general expression of the form 
\be\label{eq2}
R_{\mu\nu}-\frac{1}{2}g_{\mu\nu}R=\kappa^2\left[T_{\mu\nu}^m+T_{\mu\nu}\right]
\ee
where $\kappa^2=8\pi G$, $T_{\mu\nu}^m$ is the usual energy-momentum tensor for matter component, the tensor $T_{\mu\nu}$ represents the variation of the terms which depend on the scalar field $\phi$ and can be written as
\be\label{eq3}
T_{\mu\nu}=T_{\mu\nu}^{\phi}+T_{\mu\nu}^{\xi}+T_{\mu\nu}^{\eta}+T_{\mu\nu}^{GB}
\ee
where $T_{\mu\nu}^{\phi}$,  correspond to the variations of the standard minimally coupled terms, $T_{\mu\nu}^{\xi}$, $T_{\mu\nu}^{\eta}$ come from the $\xi$ and $\eta$ kinetic couplings respectively, and $T_{\mu\nu}^{GB}$ comes from the variation of the coupling with GB. 
Due to the kinetic coupling with curvature and the GB coupling, the quantities derived from this energy-momentum tensors will be considered as effective ones. The variations are given by
\be\label{eq4}
T_{\mu\nu}^{\phi}=\nabla_{\mu}\phi\nabla_{\nu}\phi-\frac{1}{2}g_{\mu\nu}\nabla_{\lambda}\phi\nabla^{\lambda}\phi
-g_{\mu\nu}V(\phi)
\ee
\be\label{eq5}
\begin{aligned}
T_{\mu\nu}^{\xi}=&\xi\Big[\left(R_{\mu\nu}-\frac{1}{2}g_{\mu\nu}R\right)\left(F_1(\phi)\nabla_{\lambda}\phi\nabla^{\lambda}\phi\right)+g_{\mu\nu}\nabla_{\lambda}\nabla^{\lambda}\left(F_1(\phi)\nabla_{\gamma}\phi\nabla^{\gamma}\phi\right)\\
&-\frac{1}{2}(\nabla_{\mu}\nabla_{\nu}+\nabla_{\nu}\nabla_{\mu})\left(F_1(\phi)\nabla_{\lambda}\phi\nabla^{\lambda}\phi\right)+R\left(F_1(\phi)\nabla_{\mu}\phi\nabla_{\nu}\phi\right)\Big]
\end{aligned}
\ee
\be\label{eq6}
\begin{aligned}
T_{\mu\nu}^{\eta}=&\eta\Big[F_1(\phi)\left(R_{\mu\lambda}\nabla^{\lambda}\phi\nabla_{\nu}\phi+R_{\nu\lambda}\nabla^{\lambda}\phi\nabla_{\mu}\phi\right)-\frac{1}{2}g_{\mu\nu}R_{\lambda\gamma}\left(F_1(\phi)\nabla^{\lambda}\phi\nabla^{\gamma}\phi\right)\\
&-\frac{1}{2}\left(\nabla_{\lambda}\nabla_{\mu}\left(F_1(\phi)\nabla^{\lambda}\phi\nabla_{\nu}\phi\right)+\nabla_{\lambda}\nabla_{\nu}\left(F_1(\phi)\nabla^{\lambda}\phi\nabla_{\mu}\phi\right)\right)\\
&+\frac{1}{2}\nabla_{\lambda}\nabla^{\lambda}\left(F_1(\phi)\nabla_{\mu}\phi\nabla_{\nu}\phi\right)+\frac{1}{2}g_{\mu\nu}\nabla_{\lambda}\nabla_{\gamma}\left(F_1(\phi)\nabla^{\lambda}\phi\nabla^{\gamma}\phi\right)\Big]
\end{aligned}
\ee
and 
\be\label{eq7a}
\begin{aligned}
T_{\mu\nu}^{GB}=&4\Big([\nabla_{\mu}\nabla_{\nu}F_2(\phi)]R-g_{\mu\nu}[\nabla_{\rho}\nabla^{\rho}F_2(\phi)]R-2[\nabla^{\rho}\nabla_{\mu}F_2(\phi)]R_{\nu\rho}-2[\nabla^{\rho}\nabla_{\nu}F_2(\phi)]R_{\nu\rho}\\
&+2[\nabla_{\rho}\nabla^{\rho}F_2(\phi)]R_{\mu\nu}+2g_{\mu\nu}[\nabla^{\rho}\nabla^{\sigma}F_2(\phi)]R_{\rho\sigma}-2[\nabla^{\rho}\nabla^{\sigma}F_2(\phi)]R_{\mu\rho\nu\sigma}\Big)
\end{aligned}
\ee
In this last expression the properties of the 4-dimensional GB invariant have been used (see \cite{farhoudi}, \cite{sergei12}).
Variating with respect to the scalar field gives the equation of motion
\be\label{eq7}
\begin{aligned}
&-\frac{1}{\sqrt{-g}}\partial_{\mu}\left[\sqrt{-g}\left(\xi R F_1(\phi)\partial^{\mu}\phi+\eta R^{\mu\nu}F_1(\phi)\partial_{\nu}\phi+\partial^{\mu}\phi\right)\right]+\frac{dV}{d\phi}+\\
&\frac{dF_1}{d\phi}\left(\xi R\partial_{\mu}\phi\partial^{\mu}\phi+\eta R_{\mu\nu}\partial^{\mu}\phi\partial^{\nu}\phi\right)-\frac{dF_2}{d\phi}{\cal G}=0
\end{aligned}
\ee
Considering the spatially-flat Friedmann-Robertson-Walker (FRW) metric,
\be\label{eq8}
ds^2=-dt^2+a(t)^2\left(dr^2+r^2d\Omega^2\right)
\ee
And assuming an homogeneous time-depending scalar field $\phi$ , the $(00)$ and $(11)$ components of the Eq. (\ref{eq2}), from (\ref{eq3}-\ref{eq7a}) take the form (with the Hubble parameter $H=\dot{a}/a$)
\be\label{eq9}
H^2=\frac{\kappa^2}{3}\rho_{eff}
\ee
with $\rho_{eff}$ given by
\be\label{eq9a}
\begin{aligned}
\rho_{eff}=&\Big[\frac{1}{2}\dot{\phi}^2+V(\phi)+9\xi H^2F_1(\phi)\dot{\phi}^2+3(2\xi+\eta)\dot{H}F_1(\phi)\dot{\phi}^2\\
&-3(2\xi+\eta)H F_1(\phi)\dot{\phi}\ddot{\phi}-\frac{3}{2}(2\xi+\eta)H \frac{dF_1}{d\phi}\dot{\phi}^3-24H^3\frac{dF_2}{d\phi}\dot{\phi}\Big]
\end{aligned}
\ee
and
\be\label{eq10}
-2\dot{H}-3H^2=\kappa^2 p_{eff}
\ee
with $p_{eff}$ given by
\be\label{eq10a}
\begin{aligned}
p_{eff}=&\Big[\frac{1}{2}\dot{\phi}^2-V(\phi)+3(\xi+\eta)H^2F_1(\phi)\dot{\phi}^2+2(\xi+\eta)\dot{H}F_1(\phi)\dot{\phi}^2\\
&+4(\xi+\eta)H F_1(\phi)\dot{\phi}\ddot{\phi}+2(\xi+\eta)H\frac{dF_1}{d\phi}\dot{\phi}^3\\
&+(2\xi+\eta)\left(F_1(\phi)\ddot{\phi}^2+F_1(\phi)\dot{\phi}\dddot{\phi}+\frac{5}{2}\frac{dF_1}{d\phi}\dot{\phi}^2\ddot{\phi}+\frac{1}{2}\frac{d^2F_1}{d\phi^2}\dot{\phi}^4\right)\\
&+8H^2\frac{dF_2}{d\phi}\ddot{\phi}+8H^2\frac{d^2F_2}{d\phi^2}\dot{\phi}^2+16H\dot{H}\frac{dF_2}{d\phi}\dot{\phi}+16H^3\frac{dF_2}{d\phi}\dot{\phi}\Big]
\end{aligned}
\ee
where we have assumed scalar field dominance (i.e. $T^m_{\mu\nu}=0$). The equation of motion for the scalar field (\ref{eq7}) takes the form
\be\label{eq11}
\begin{aligned}
&\ddot{\phi}+3H\dot{\phi}+\frac{dV}{d\phi}+3(2\xi+\eta)\ddot{H}F_1(\phi)\dot{\phi}+
3(14\xi+5\eta)H\dot{H}F_1(\phi)\dot{\phi}\\
&+\frac{3}{2}(2\xi+\eta)\dot{H}\left(2F_1(\phi)\ddot{\phi}+\frac{dF_1}{d\phi}\dot{\phi}^2\right)+
\frac{3}{2}(4\xi+\eta)H^2\left(2F_1(\phi)\ddot{\phi}+\frac{dF_1}{d\phi}\dot{\phi}^2\right)\\
&+9(4\xi+\eta)H^3F_1(\phi)\dot{\phi}-24\left(\dot{H}H^2+H^4\right)\frac{dF_2}{d\phi}=0
\end{aligned}
\ee
where the first three terms correspond to the minimally coupled field. In what follows we study the cosmological consequences of this equations, under some conditions that simplify the search for solutions.\\
\noindent An important simplification of the Eqs. (\ref{eq9}-\ref{eq11}) takes place, under the restriction on $\xi$ and $\eta$ given by 
\be\label{eq1a}
\eta+2\xi=0
\ee
Under this restriction all the kinetic couplings that appear in the action (\ref{eq1}) become summarized in the term $G_{\mu\nu}\partial^{\mu}\phi\partial^{\nu}\phi$, where $G_{\mu\nu}=R_{\mu\nu}-\frac{1}{2}g_{\mu\nu}R$. In this case the field equations (\ref{eq9}-\ref{eq11}) contain only second derivatives of the metric and the scalar field, avoiding problems with higher order derivatives \cite{capozziello1, sushkov}. The modified Friedmann equations (\ref{eq9}) and (\ref{eq10}) take the form
\be\label{eq12}
H^2=\frac{\kappa^2}{3}\left(\frac{1}{2}\dot{\phi}^2+V(\phi)+9\xi H^2F_1(\phi)\dot{\phi}^2-24H^3\frac{dF_2}{d\phi}\dot{\phi}\right)
\ee
and
\be\label{eq13}
\begin{aligned}
-2\dot{H}-3H^2=&\kappa^2\Big[\frac{1}{2}\dot{\phi}^2-V(\phi)-\xi\left(3H^2+2\dot{H}\right)F_1(\phi)\dot{\phi}^2-2\xi H\left(2F_1(\phi)\dot{\phi}\ddot{\phi}+\frac{dF_1}{d\phi}\dot{\phi}^3\right)\\
&+8H^2\frac{dF_2}{d\phi}\ddot{\phi}+8H^2\frac{d^2F_2}{d\phi^2}\dot{\phi}^2+16H\dot{H}\frac{dF_2}{d\phi}\dot{\phi}+16H^3\frac{dF_2}{d\phi}\dot{\phi}\Big]
\end{aligned}
\ee
The equation of motion reduces to
\be\label{eq14}
\begin{aligned}
&\ddot{\phi}+3H\dot{\phi}+\frac{dV}{d\phi}+3\xi H^2\left(2F(\phi)\ddot{\phi}+\frac{dF}{d\phi}\dot{\phi}^2\right)
+18\xi H^3F(\phi)\dot{\phi}+\\
&12\xi H\dot{H}F(\phi)\dot{\phi}-24\left(\dot{H}H^2+H^4\right)\frac{dF_2}{d\phi}=0
\end{aligned}
\ee
\noindent Note that, independently of $F_1(\phi)$, and for $F_2(\phi)=const.$, assuming an asymptotic behavior of the scalar field as $\phi=\phi_0=const.$, gives rise to de Sitter solution, as can be seen from Eqs. (\ref{eq9}) and (\ref{eq11}). From (\ref{eq9}) and (\ref{eq11}) it follows that $V=V_0=const$ and $H=H_0=\kappa\sqrt{V_0/3}$.\\
Next we try to study cosmological solutions of this model, giving rise to to accelerated expansion and acceptable behavior of the equation of state parameter (EoS). 


\section{Power law solutions}
It is of interest to derive solutions that
give rise to a power-law expansion ($a\propto t^p$), as it's well known that this kind of expansion is characteristic of the evolution at the early stages of radiation and matter dominance, and also may respond for accelerated expansion.  We will consider the effects of this new kinetic coupling in the cosmological dynamics, in the case of scalar field dominance (we further neglect any background radiation or matter contribution). 
\subsection{Power law with free kinetic term}
In this section we will consider the model with the coupling $F_1(\phi)=1/\phi^2$. This function has the important property of leaving dimensionless the coupling constant $\xi$ in (\ref{eq12}).\\
Replacing $dF_2/dt$ From (\ref{eq12}) into (\ref{eq14}), and using $F_1(\phi)=1/\phi^2$, yields
\be\label{eq14a}
\begin{aligned}
&\frac{H}{2}\frac{d}{dt}(\dot{\phi}^2)+(5H^2-\dot{H})\frac{\phi^2}{2}+3\xi H^3\frac{d}{dt}\left(\frac{\dot{\phi}^2}{\phi^2}\right)+9\xi H^4\frac{\dot{\phi}^2}{\phi^2}+3\xi H^2\dot{H}\frac{\dot{\phi}^2}{\phi^2}\\
&+H\frac{dV}{dt}-\left(H^2+\dot{H}\right)V+\frac{3}{\kappa^2}H^2\left(H^2+\dot{H}\right)=0
\end{aligned}
\ee
where we have multiplied Eq. (\ref{eq14}) by $\dot{\phi}$ and used $\dot{\phi}d/d\phi=d/dt$. Assuming the following time dependence for the scalar field and the Hubble function
\be\label{eq14b}
\phi=\frac{\sqrt{2}}{t},\,\,\,\,\,\, H=\frac{p}{t}
\ee
and integrating (\ref{eq14a}) we find the potential
\be\label{eq14c}
V(\phi)=\frac{9\xi p^3-9\xi p^2+5p-3}{4(p+3)}\phi^4+\frac{3p^2(p-1)}{2(p+1)\kappa^2}\phi^2+\frac{2^{(p-1)/2}C}{\phi^{p-1}}
\ee
where we have replaced $t=\sqrt{2}/\phi$. Solving the Eq. (\ref{eq12}) with respect to $F_2$, we find
\be\label{eq14d}
F_2(\phi)=-\frac{3\xi p^2+3\xi p+1}{4p^2(p+3)}\ln\left(\frac{\phi}{\sqrt{2}}\right)-\frac{1}{4p(p+1)\kappa^2}\frac{1}{\phi^2}+\frac{2^{(p+3)/2}C}{24p^3(p+3)}\frac{1}{\phi^{p+3}}
\ee
This solution gives rise to accelerated expansion, provided $p>1$. A remarkable property of the potential (\ref{eq14c}), is that for the particular choice $C=0$, becomes of the Higgs type.\\
Let's consider the particular case of the model (\ref{eq1}) without free kinetic term. In this case we are in the frames of the models (\cite{sergei,allemandi,sergei1,sergei2}). This also applies if the the slow-roll condition $\dot{\phi}^2<<V(\phi)$ is considered, which can take place in the contexts of dark energy or inflationary cosmology.
\subsection{Derivative coupling with $V=0$}
Let us then, begin with the simple case of strictly non-minimal kinetic coupling, without potential term. In this simple case we may consider the general Eqs. (\ref{eq9}) and (\ref{eq11}), which after eliminating the free kinetic term an potential become
\be\label{eq15}
H^2=\frac{\kappa^2}{3}\Big[9\xi H^2F_1(\phi)\dot{\phi}^2+3(2\xi+\eta)\dot{H}F_1(\phi)\dot{\phi}^2
-\frac{3}{2}(2\xi+\eta)H \frac{d}{dt}(F_1\dot{\phi}^2)-24H^3\frac{dF_2}{dt}\Big]
\ee
and 
\be\label{eq16}
\begin{aligned}
&3(2\xi+\eta)\ddot{H}F_1(\phi)\dot{\phi}^2+3(14\xi+5\eta)H\dot{H}F_1(\phi)\dot{\phi}^2+\frac{3}{2}(2\xi+\eta)\dot{H}\frac{d}{dt}(F_1\dot{\phi}^2)+\\
&\frac{3}{2}(4\xi+\eta)H^2\frac{d}{dt}(F_1\dot{\phi}^2)+9(4\xi+\eta)H^3F_1(\phi)\dot{\phi}^2-24\left(\dot{H}H^2+H^4\right)\frac{dF_2}{dt}=0
\end{aligned}
\ee
where we have multiplied the Eq. (\ref{eq11}) by $\dot{\phi}$. Let's propose now a power law solution $H=p/t$, and the coupling $F_1$ satisfying $F_1\dot{\phi}^2=C_1$. Then replacing in Eq. (\ref{eq16})we obtain
\be\label{eq17}
\frac{dF_2}{dt}=C_2t,\,\,\,\,\,\,\, C_2=\frac{\left(4\xi+2\eta-(14\xi+5\eta)p+3(4\xi+\eta)p^2\right)C_1}{8p(p-1)}
\ee
Integrating this equation gives 
\be\label{eq18}
F_2(t)=\frac{1}{2}C_2 t^2
\ee
where we equated the integration constant to zero. Replacing this solution together with $H=p/t$ and $F_1\dot{\phi}^2=C_1$ in Eq. (\ref{eq15}), we obtain the following relation between $C_1$ and $p$ (setting $\kappa^2=0$)
\be\label{eq19}
C_1=\frac{(p-1)p}{(3p-1)(2\xi+\eta-3\xi p-\eta p})
\ee
giving rise to accelerated expansion provided $p>1$. Thus for example taking $F_1=F_0e^{2\phi/\phi_0}$ and limiting to those values of $C_1>0$, the scalar field behaves as (from $F_1\dot{\phi}^2=C_1$)
\be\label{eq20}
\phi=\phi_0 \ln\left(\sqrt{\frac{C_1}{F_0}}\frac{t}{\phi_0}\right)
\ee
and the GB coupling function $F_2$ from (\ref{eq18})  becomes 
\be\label{eq21}
F_2=\frac{C_2\phi_0^2F_0}{2C_1}e^{2\phi/\phi_0}
\ee
Under the restriction (\ref{eq1a}), the relation between $C_1$, $C_2$ and $p$ becomes $C_1=\frac{1-p}{(3p-1)\xi}$ and $C_2=\frac{2-3p}{12p-4}$, which also gives rise to accelerated expansion for $p>1$. For the same restriction (\ref{eq1a}) in the model without GB coupling, the power-law solution  was limited to $p=2/3$ \cite{granda1}.\\
\subsection{Derivative coupling with potential}
From now on we will consider the restriction (\ref{eq1a}). We will look for the shape of the potential $V(\phi)$  and couplings corresponding to the asymptotic behavior giving rise to power-law accelerated expansion. Dropping the free kinetic terms in (\ref{eq12}) and (\ref{eq14}) and replacing $H=p/t$ we obtain
\be\label{eq22}
\frac{3}{\kappa^2}\frac{p^2}{t^2}=V+9\xi\frac{p^2}{t^2}(F_1\dot{\phi}^2)-24\frac{p^3}{t^3}\frac{dF_2}{dt}
\ee
and 
\be\label{eq23}
\frac{dV}{dt}+3\xi\frac{p^2}{t^2}\frac{d}{dt}\left(F_1\dot{\phi}^2\right)+18\xi\frac{p^3}{t^3}\left(F_1\dot{\phi}^2\right)-12\xi\frac{p^2}{t^3}\left(F_1\dot{\phi}^2\right)-24\frac{p^3(p-1)}{t^4}\frac{dF_2}{dt}=0
\ee
where in this last equation we have multiplied by $\dot{\phi}$ and the first two terms of Eq. (\ref{eq14}) have been dropped due to the absence of the free kinetic term.
using the additional freedom provided by the GB coupling we can impose the following restriction on the potential
\be\label{eq24}
\frac{dV}{dt}-24\frac{p^3(p-1)}{t^4}\frac{dF_2}{dt}=0
\ee
which converts Eq. (\ref{eq23}) into a first order differential equation with respect to the variable $(F_1\dot{\phi}^2)$:
\be\label{eq24a}
3\xi\frac{p^2}{t^2}\frac{d}{dt}\left(F_1\dot{\phi}^2\right)+18\xi\frac{p^3}{t^3}\left(F_1\dot{\phi}^2\right)-12\xi\frac{p^2}{t^3}\left(F_1\dot{\phi}^2\right)
\ee
Integrating this equation with respect to $(F_1\dot{\phi}^2)$, gives
\be\label{eq25}
F_1\dot{\phi}^2=\chi_0\left(\frac{t}{t_0}\right)^{-(6p-4)}
\ee
using the restriction (\ref{eq24}) in Eq. (\ref{eq22}), we get the following differential equation for the potential
\be\label{eq26}
\frac{t}{p-1}\frac{dV}{dt}-V+\frac{3p^2}{\kappa^2t^2}-\frac{9\xi\chi_0 p^2 t_0^{6p-4}}{t^{6p-2}}=0
\ee
Integrating this equation gives
\be\label{eq27}
V=C t^{p-1}+\frac{3(p-1)p^2}{(p+1)\kappa^2}\frac{1}{t^2}+\frac{9\xi\chi_0(p-1)p^3t_0^{6p-4}}{3-7p}\frac{1}{t^{6p-2}}
\ee
and from (\ref{eq24}) we find the expression for the GB coupling
\be\label{eq28}
F_2=\frac{C}{24p^3(p+3)}t^{p+3}-\frac{3p^2}{(p+1)\kappa^2}t^2+\frac{3p^3(3p-1)\xi\chi_0 t_0^{6p-4}}{(7p-3)(1-p)}t^{6-6p}
\ee
Finally, from Eq. (\ref{eq25}), for a given $F_1$ we can find the scalar field as a function of time and then express the potential $V$ and the coupling $F_2$ in terms of the scalar field. But the expression (\ref{eq25}) also allows to propose the scalar field as function of time, and find the expression for the kinetic coupling $F_1$. Considering both possibilities, we first propose the kinetic coupling as $F_1=F_0e^{-2\phi/\phi_0}$, which from $(\ref{eq25})$ gives (up to additive constant)
\be\label{eq29}
\phi=3(p-1)\phi_0 \ln\left(\frac{t}{t_0}\right)
\ee
making the rescaling $3(p-1)\phi_0\rightarrow \phi_0$ and replacing $t$ in Eqs. (\ref{eq27}) and (\ref{eq28}) we find the following expressions for the potential and GB coupling
\be\label{eq30}
V=C t_0^{p-1}e^{(p-1)\phi/\phi_0}+\frac{3(p-1)p^2}{(p+1)\kappa^2t_0^2}e^{-2\phi/\phi_0}+\frac{9\xi\chi_0(p-1)p^3}{(3-7p)t_0^2}e^{-2(3p-1)\phi/\phi_0}
\ee
and
\be\label{eq31}
F_2=\frac{Ct_0^{p+3}}{24p^3(p+3)}e^{(p+3)\phi/\phi_0}-\frac{3p^2t_0^2}{(p+1)\kappa^2}e^{2\phi/\phi_0}+\frac{3p^3(3p-1)\xi\chi_0 t_0^2}{(7p-3)(1-p)}e^{-6(p-1)\phi/\phi_0}
\ee
By assuming now for the scalar field the simple expression $\phi=t^{-1}$, from (\ref{eq25}) it follows 
\be\label{eq32}
F_1=\frac{\chi_0 t_0^{6p-4}}{t^{6p-8}}=\chi_0 t_0^{6p-4}\phi^{6p-8}
\ee
and the potential $V$ and GB coupling $F_2$ are obtained from (\ref{eq27}) and (\ref{eq28}) by making the replacement $t=1/\phi$. In this case the GB coupling and the scalar potential are combinations of different powers of the scalar field. This shows that in the case when the free kinetic term is neglected, we always can find the appropriate couplings and potential giving rise to power-law expansion. Note that looking at the time dependence in all considered solutions, the GB coupling decays faster than the coupling $F_2$ grows or decays (making the integration constant $C=0$ in (\ref{eq14d}) and (\ref{eq28})). Then the term $F_2{\cal G}$ decays with time, behaving properly in the current low-curvature universe.
\section{Dynamically varying equation of state}
So far we have considered solutions of the power law type, which are characterized by a constant EoS parameter. Power law solutions may describe a decelerating universe with matter or radiation dominance, as well as an accelerating universe, but cannot describe the transition from one phase to another, due to the non-dynamical nature of the EoS parameter. In the present model one can explicitly construct a solution which admits the transition from the matter-dominant period to the acceleration phase dominated by dark energy. Considering as before the coupling $F_1=1/\phi^2$, the scalar field $\phi=\sqrt{2}/t$, and proposing the Hubble function
\be\label{eq32a}
H(t)=\gamma+\frac{p}{t},
\ee
with $\gamma$ and $p$ constants, the equation (\ref{eq14a}) takes the form (setting $\kappa^2=1$)
\be\label{eq32b}
\begin{aligned}
&\left(pt^5+\gamma t^6\right)\frac{dV}{dt}+\left(p(1-p)t^4-2p\gamma t^5-\gamma^2 t^6\right)V+3\gamma^4 t^6+12p\gamma^3 t^5\\
&+3\gamma^2\left(6p^2-p+3\xi\gamma^2\right)t^4+6\gamma\left(2p^3-p^2-\xi\gamma^2+6p\xi\gamma^2\right)t^3\\
&+\left(3p^4-3p^3+5\gamma^2-21p\xi\gamma^2+54p^2\xi\gamma^2\right)t^2
+2\gamma\left(18p^3\xi-5p-2\right)t\\
&+9\xi p^4-9\xi p^3+5p^2-3p=0
\end{aligned}
\ee
To solve this equation we propose a particular solution of the form
\be\label{eq32c}
V(t)=V_0+\frac{A}{t}+\frac{B}{t^2}+\frac{C}{t^3}+\frac{D}{t^4}
\ee
Replacing this solution in (\ref{eq32b}) and solving, we find two solutions:
\be\label{eq32d}
V_0=0,\,\,\,\,\, A=0, \,\,\,\,\, B=\frac{3p^2(p-1)}{p+1},\,\,\,\,\, C=0,\,\,\,\,\,\, D=\frac{9\xi p^3-9\xi p^2+5p-3}{p+3},\,\,\,\, \gamma=0
\ee
which reproduces the previous result for power law expansion (see (\ref{eq14c}) with $C=0$), and
\be\label{eq32e}
\begin{aligned}
&V_0=3\gamma^2,\,\,\,\,\, \xi=-\frac{6p+5p^2+p^3-\gamma^2}{2(p-6)\gamma^2},\,\,\,\, A=6p\gamma\\ &B=\frac{3\left(3\gamma^2+6p-31p^2-p^3\right)}{2(p-6)},\,\,\,\, C=-\frac{3\left(3p^4+11p^3-6p^2-3p\gamma^2+4\gamma^2\right)}{(p-6)\gamma}\\
&D=\frac{12\gamma^2-26p\gamma^2+9p^2\gamma^2+18p^3-9p^4-9p^5}{2(p-6)\gamma^2}
\end{aligned}
\ee
The potential can be written explicitly in terms of the scalar field, using $t=\sqrt{2}/\phi$
\be\label{eq32f} 
V(\phi)=V_0+\frac{A}{\sqrt{2}}\phi+\frac{B}{2}\phi^2+\frac{C}{2\sqrt{2}}\phi^3+\frac{D}{4}\phi^4
\ee
replacing $V$ from (\ref{eq32c}) in Eq. (\ref{eq12}), using (\ref{eq32e}) and solving for the GB coupling $F_2$, we find
\be\label{eq32g}
\begin{aligned}
F_2(\phi)=&\frac{(\gamma^2-2p^3)\phi}{2p(6-p)\gamma^2(p\phi+\sqrt{2}\gamma)}+\\
&\frac{\left(3p^4+9p^3+6p^2+(4-3p)\gamma^2\right)\left(\ln \sqrt{2}-\ln(p\phi+\sqrt{2}\gamma)\right)}{8p^2(6-p)\gamma^2}+C
\end{aligned}
\ee
where after solving with respect to time variable, we have replaced $t$ by $\sqrt{2}/\phi$. 
This expressions are valid for any $p\neq 6$, and therefore the model may reproduce known cosmological scenarios. As follows from (\ref{eq32a}) the EoS parameter is
\be\label{eq32h}
w=-1+\frac{2}{3}\frac{p}{\gamma t^2+2\gamma pt+p^2}
\ee
for small $t$ we can approximate $w\approx -1+2/(3p)$, and the second term in (\ref{eq32a}) dominates, giving a scale factor $a\sim t^p$. Therefore, if $p=2/3$ or $p=1/2$ the matter dominated or radiation dominated period could be realized. At late times, the first term $\gamma$ in (\ref{eq32a}) dominates and the universe evolves towards asymptotically de Sitter space time. Hence, the model is consistent with current phenomenology.\\

Let's consider now the model without free kinetic term, and kinetic coupling $F_1(\phi)=1/\phi^2$. Working with the variable $x=\ln a$, the Eqs. (\ref{eq12}) and (\ref{eq14}) (after multiplying by $\dot{\phi}$) take the form
\be\label{eq33}
\frac{3H^2}{\kappa^2}=V+9\xi H^4 \chi^2-24H^4\frac{dF_2}{dx},\,\,\,\,\,\,\, \chi=\frac{1}{\phi}\frac{d\phi}{dx}
\ee
and
\be\label{eq34}
\frac{dV}{dx}+9\xi H^2\frac{dH^2}{dx}\chi^2+3\xi H^4\frac{d\chi^2}{dx}+18\xi H^4\chi^2-12H^2\left(\dot{H}+H^2\right)\frac{dF_2}{dx}=0
\ee
where we used $\frac{d}{dt}=H\frac{d}{dx}$.\\
Considering the particular form of the scalar field $\phi=\phi_0 e^{-\alpha x}$ and the GB coupling $F_2=\lambda\ln(\phi/\phi_0)$, $\chi$ takes the value $\chi=-\alpha$ and $dF_2/dx=\lambda\chi=-\lambda\alpha$. Then, taking the derivative of Eq. (\ref{eq33}) and replacing $dV/dx$ in (\ref{eq34}) we obtain the following evolution equation for the Hubble parameter
\be\label{eq35}
\frac{1}{\kappa^2}\frac{dH^2}{dx}-3\alpha\left(\xi\alpha+4\lambda\right)H^2\frac{dH^2}{dx}+2\alpha\left(3\xi\alpha+4\lambda\right)H^4=0
\ee
the solution to this equation is given by (setting $\kappa^2=1$)
\be\label{eq36}
H^2(x)=-\left(3\alpha(\xi\alpha+4\lambda) W\left[-\frac{1}{3\alpha(\alpha\xi+4\lambda)}e^{-\frac{2(3\alpha\xi+4\lambda)}{3(\alpha\xi+4\lambda)}x-\frac{C}{3\alpha(\xi\alpha+4\lambda)}}\right]\right)^{-1}
\ee
where $W$ is the Lambert $W$ function and $C$ is the integration constant. From (\ref{eq36}) and the properties of the $W$ function, it follows that for $\alpha$ and $\xi$ satisfying the inequality $\alpha(\alpha\xi+4\lambda)<0$, $H^2$ will always positive (otherwise we may enter in a region where $W$ becomes complex, see \cite{granda} for details). Note that for $\lambda=0$ we obtain the solution described in \cite{granda}, and $\lambda=-3\alpha\xi/4$ gives rise to de Sitter solution. Choosing $\lambda=1$ and in the redshift variable $z$, the Hubble parameter takes the form
\be\label{eq37}
H^2(x)=-\left(3\alpha(\xi\alpha+4) W\left[-\frac{e^{-\frac{C}{3\alpha(\xi\alpha+4)}}}{3\alpha(\alpha\xi+4)}(1+z)^{\frac{2(3\alpha\xi+4)}{3(\alpha\xi+4)}}\right]\right)^{-1}
\ee
and evaluating the equation of state (EoS) parameter $w_{eff}(z)=-1+\frac{1+z}{3H^2}\frac{dH^2}{dz}$, from (\ref{eq37}) one obtains
\be\label{eq38}
w=-1-\frac{8+6\eta}{9(\alpha\xi+4)\left[1+W\left[-\frac{e^{-\frac{C}{3\alpha(\eta+4)}}}{3\alpha(\eta+4)}(1+z)^{\frac{2(3\eta+4)}{3(\eta+4)}}\right]\right]}
\ee
where $\eta=\alpha\xi$. In order to keep $H^2$ positive, $\alpha$ and $\eta$ should satisfy $\alpha(\eta+4)<0$. From the expression (\ref{eq38}) it follows that the behavior of 
$w$ at the limits $z\rightarrow\infty$ and $z\rightarrow -1$ does not depend on the values of $\alpha$ and $C$. Thus, $\alpha$ and $C$ can be used to have an appropriate value of 
the current EoS parameter $w_0$. An interesting value for $\eta$ is $\eta=-28/9$, which gives a radiation dominated universe at very early times $w\big|_{z\rightarrow\infty}=1/3$ and cosmological constant solution at future $w\big|_{z\rightarrow -1}=-1$, independently of the values of $\alpha$ and $C$. Taking $\alpha=-1$ and $C=200$ the current value of the EoS is $w_0\approx-0.98$. Another behavior takes place for $\eta=-24/9$. In this case the EoS describes an effective quintessence evolution, as it starts in the lower limit of the accelerated phase $w\big|_{z\rightarrow\infty}=-1/3$ and ends in a de Sitter phase $w\big|_{z\rightarrow -1}=-1$. And the current value of the EoS is $w_0\approx-0.97$ with $\alpha=-1$ and $C=100$. An universe evolving from a matter dominated phase at high redshifts $w\big|_{z\rightarrow\infty}=0$, is obtained for $\eta=-44/15$. This solution evolves towards de Sitter phase at far future, with an EoS at the present  $w_0\approx-0.99$ (for $\alpha=-0.1$ and $C=50$). There is also a possibility for a universe in the phantom phase, with an EoS evolving from $w=-1$ at $z\rightarrow \infty$ to $w<-1$ at $z\rightarrow -1$ (i.e. starting in a de Sitter phase and ending in phantom phase) with an EoS currently less than $-1$. Thus, for instance $\eta=1$, gives $w=-1$ at $z\rightarrow \infty$ and $w=-59/45$ at $z\rightarrow -1$, with the current value $w_0=-31/24$ for $\alpha=-1$ and $C=1$, or taking $C=50$ we have $w_0\sim -1.17$. In this case the model describes pure dark energy behavior.\\ 
\noindent From the expression for $H^2$ (\ref{eq37}), it follows that the effective energy density given by $\rho_{eff}=3H^2/\kappa^2$ shows two types of behaviors: for a solution starting from radiation or matter dominance, or purely quintessence behavior (i.e. $\eta=-28/9$,$-44/15$,$-24/9$), the evolution begins from Big Bang singularity, as  $\rho_{eff}\big|_{z\rightarrow\infty}=\infty$ and presents no future singularities, $\rho_{eff}\big|_{z\rightarrow -1}=0$ . And for solutions describing a phantom phase (i.e. $\eta=1$), $\rho_{eff}$ tends to zero at high redshifts, validating the assumption that the dark energy was negligible at early times, but presents asymptotical singularity ($\rho_{eff}\big|_{z\rightarrow -1}=\infty$), characteristic of the phantom behavior.\\
Another interesting solution takes place if we impose the restriction $\xi\alpha+4\lambda=0$, in Eq. (\ref{eq35}), which after integration gives the result (setting $\kappa^2=1$)
\be\label{eq39}
H^2=\frac{1}{4\xi\alpha^2 x+C}=\frac{1}{4\xi\alpha^2\ln a+C}
\ee
where $C$ is the integration constant and we replaced $x=\ln a$. Integrating this equation with respect to time, gives the scale factor
\be\label{eq40}
a(t)=a_0 \exp\left[\left(\frac{9}{16\xi\alpha^2}\right)^{1/3}t^{2/3}\right]
\ee
hence, when $t$ is small we may expand (\ref{eq40}) keeping the first two terms, and the scale factor behaves as $a\sim t^{2/3}$. Therefore, the matter-dominated period could be realized. On the other hand, when $t$ is large, from (\ref{eq40}) the Hubble function becomes: $H=\frac{(6\xi\alpha^2)^{-1/3}}{t^{1/3}}$. Then $w_{eff}=-1+\frac{2(6\xi\alpha^2)^{1/3}}{9t^{2/3}}$, and therefore the universe accelerates and is asymptotically de Sitter space. 
\section{Discussion}
We have considered a model of scalar field that contains non-minimally kinetic terms coupled to curvature and the Gauss Bonnet four dimensional invariant coupled to the scalar field. This GB coupling has the advantage that does not make contributions higher than second order (in the metric) to the equations of motion, and therefore does not introduce ghost terms into the theory. Nevertheless, this new coupling introduces a new degree of freedom that could be exploited to enrich the number of possible evolutionary scenarios that explain the nature of the dark energy.
Power law solutions have been found for different configurations of the model, namely, the complete model, and the model without free kinetic term, with and without potential. Under the restriction $\eta+2\xi=0$, and neglecting the free kinetic term, the GB coupling allows power-law solution with accelerated expansion ($p>1$), improving the results in absence of GB coupling where the only decelerated expansion with fixed $p=2/3$ was possible \cite{granda}. The power-law solutions with ($p>1$) are also possible in absence of potential, or with potential and GB coupling as a combination of exponentials of the scalar field.\\
Studying cosmological solutions with varying EoS, we considered two cases. In the complete model we found an interesting solution that contains the two asymptotic behaviors corresponding to early time (matter or radiation dominated universe) and late time dark energy dominated universe, evolving towards the de Sitter phase \cite{sergei11}. In absence of free kinetic term, we considered the couplings
$F_1=1/\phi^2$ (which leaves the coupling constant $\xi$ dimensionless), $F_2=\lambda\ln(\phi/\phi_0)$, and assumed the scalar field $\phi=\phi_0 e^{-\alpha x}$ (which is a reasonable choice as $e^{-\alpha x}=(1+z)^{\alpha}$), we obtained a variety of solutions with different asymptotical desirable behaviors, and that reproduce the current observed value of $w$. 
The solutions with $\eta=-44/15, -28/9$ describe an universe starting asymptotically in matter or radiation dominance, and evolving towards the future de Sitter solution, with current values $w_0\sim -0.99, -0.98$ respectively. Therefore this solutions present transition between decelerated and accelerated phases. Taking for instance $\eta=-24/9$ gives rise to an universe in quintessential phase starting with $w=-1/3$, with current $w_0\sim -0.97$ and ending in a de Sitter phase with $w=-1$. There is also a possibility for a solution with universe in essentially phantom phase for $\eta=1$ beginning 
in de Sitter phase, with current $w_0=-31/24$, and evolving towards a state with $w=-59/45$ (note that in all cases the current $w_0$ may vary, depending on the choice of $\alpha$ and $C$).
This last two types of solutions belong to purely dark energy description and need the addition of matter content to achieve a phase transition between decelerated and accelerated phases. All these 
results are based on the properties of the Lambert $W$ function, which varies very slowly in an infinite (redshift) interval. These results, that were obtained thanks to the GB coupling, generalizes the one obtained in \cite{granda} as in that case the argument of the  Lambert $W$ function depended on fixed power of the redshift ($(1+z)^2$) giving only one possible cosmological scenario.\\
A remarkable property of the solution (\ref{eq36}) is that the cosmic evolution was obtained (i.e. was not proposed a priori) as a result of solving the dynamical equations for given couplings $F_1(\phi)$, $F_2(\phi)$ and scalar field $\phi$ (instead of the potential $V$), reflecting in some sense predictive power of the model, as the obtained solution is compatible with current DE phenomenology. In the DE models, mainly due to the complexity of the involved equations, one usually gives a dynamical evolution a priori, and reconstruct the model satisfying this solution. 
Is important to note, that in all solutions considered here, the Gauss-Bonnet invariant ${\cal G}$ decays faster than the coupling $F_2(\phi)$
grows or decays, so that at late times the term $F_2(\phi){\cal G}$ decays and therefore, does not substantially affect the current restrictions on the time variation of the gravitational coupling with respect to that evaluated with the kinetic coupling \cite{granda,granda1}.\\
\noindent In conclusion, we have shown that the combined effect of non minimal kinetic coupling to curvature and the Gauss Bonnet coupling to scalar field, could represent an interesting source of dark energy, which might play an important role in the explanation of the current (late time) cosmological dynamics. We have concentrated mainly on the scalar kinetic coupling coupling $F_1=1/\phi^2$, while the study of other type of couplings and their role in the late time cosmology will be considered elsewhere.


\end{document}